\documentclass[twocolumn,showpacs,preprintnumbers,
amsmath,amssymb]{revtex4}
\usepackage{graphicx}
\usepackage{dcolumn}
\usepackage{bm}
\usepackage{tensor}

\begin{document}
\title{Configuration of vortex-antivortex lattices \\
at output mirror of wide-area microchip laser}
\author{A.Yu.Okulov}
\email{alexey.okulov@gmail.com}
\homepage{https://sites.google.com/site/okulovalexey}
\affiliation{Russian Academy of Sciences, 119991, Moscow, Russia}

\date{\ November 2, 2018}

\begin{abstract}
{Square vortex lattices predicted theoretically and 
experimentally observed in diode-pumped solid state 
microchip lasers are shown to have a remarkable symmetry. 
These lattices are formed by counter-rotating vortices.
The interaction of vortices with nonlinear gain medium 
leads to precession of vortices and collective excitations 
with dynamics identical to acoustical and optical vibrations 
of atomic lattices.}
\end{abstract}


\maketitle

\section{Introduction}  
Pattern formation outside the 
equilibrium demonstrates a 
rich variety of emerging structures 
\cite{Cross:1993,Okulov:1986}.  
The vortex lattices in high Fresnel number solid-state
microchip lasers
arising within a broad range of experimental parameters 
\cite{Chen:2001} were predicted 
theoretically in \cite{Staliunas:1995}. The subsequent numerical
investigation \cite{Okulov:2008} have shown 
the deep similarity of optical vortex lattices in lasers
with spatial 
patterns that spontaneously occur in Faraday
instability \cite{Francois:2014}. Trajectories of 
individual particles of ensemble in Faraday convection 
are chaotic but the averaged mass flow 
in Faraday square cells manifests
itself as vortex lattice.  The elementary cell consists 
from a four vortices. Each 
pair of whirls located at the corners of the cell are
co-rotating and each pair of adjacent whirls at 
the side of the cell  are counter-rotating. 
The corresponding field of velocities averaged 
over ensemble of particles is continuous though each 
given particle in ensemble 
moves on complex trajectory \cite{Francois:2014}. 

The same patterns were observed in wide area 
diode-pumped solid state microchip lasers \cite{Chen:2003}. 
The Fresnel number $N_f$ in these experiments
had been varied from 10 to 1000, 
the number of synchronously counter-rotating vortices 
along  the side of the square 
was about $N_L=30$. The spatially periodic 
vortex-antivortex patterns recorded experimentally 
in a broad range of system parameters are in remarkable 
difference from conventional sets of Gaussian-Hermite ($HG$) 
or Gaussian-Laguerre ($LG$) eigenmodes. The square pattern 
occurs due to phase-locking of $HG$ or $LG$ modes via
nonlinearity inherent to gain medium \cite{Okulov:2004}. 
The theoretical study in \cite{Staliunas:1995} and subsequent 
experimental confirmation \cite{Chen:2001} overlooked the 
phase structure of optical field 
$\theta(\vec r)= arg [E_n(\vec r)]$. 
The investigation of spatial phase distribution of 
optical field and associated probability flow 
with appropriate  field of velocities 
$\vec v(\vec r) = \hbar \nabla \theta(\vec r) /m$ 
given by Madelung transform have 
shown the existence of regular array of whirls with 
opposite circulations \cite {Okulov:2009_vav}. Current 
survey is aimed to study nonstationary regimes of 
vortex-antivortex lattices formation when stationary 
solutions are perturbed by noise sources due to 
spontaneous emission $\delta E_n (\vec r)$ 
and fluctuations of optical pump $\delta N_n (\vec r)$. 
\section{Master equation and observed dynamics}
The master equation for the square lattice 
pattern formation is of
Ginzburg-Landau type \cite{Malomed:2018}. 
The other model relevant to laser cavity under
consideration is 
equivalent to Fox-Lee integral with nonlinear gain
included in explicit form \cite{Okulov:1988}:  
\begin{equation}
\label{nonlmap3}
\ E_{n+1}(\vec{r} )= \int\int^\infty_
{-\infty}\ K(\vec{r}-\vec{r'}) f(E_{n}(\vec{r'})) d^2 
\vec{r'},
\end{equation}
\begin{equation}
\label{kernelfabry3}
\ K(\vec{r}-\vec{r'} )= \frac{ikD(\vec{r'})}
{2 \pi {{L_r}} }   {\:}  exp{\:}  
[ik(\vec{r}-\vec{r'})^2 / 2{{L_r}}].
\end{equation}

\begin{eqnarray}
\label{nonlin}
f(E_{n}(\vec{r})) = 
\frac{\sigma L_a N_{n}(\vec{r}) E_{n}(\vec{r}) 
(1+i{\delta \omega}T_2) }{2}+ 
&& \nonumber \\
E_{n}(\vec{r})
+ \delta E_n (\vec r) ,
\end{eqnarray}

inversion dynamics is given by:

\begin{eqnarray}
\label{kernelfabry2}
\ N_{n+1}(\vec{r}) = N_{n}(\vec{r})  +[{\frac{N_{0}
(\vec{r})-N_{n}(\vec{r})+\delta N_n (\vec r)) }
{T_1}}-
&& \nonumber \\
{\sigma} N_{n}(\vec{r}) {\:} 
c{\:}   \epsilon_0  {\:}  |{E_{n}}|^2 / 
{ \hbar {\omega }} ]{\frac{2 L_{r}}{c}},
\end{eqnarray}

where $E_n(\vec r)$ is electric field envelope at output
mirror, $f$ is nonlinear gain transfer function,
 $L_r$ is cavity length,
$L_a<<L_r$ is a thickness of amplifying medium relevant to 
experimental situation \cite {Khazanov:2013}, 
$R$ is reflectivity of 
output mirror, $D(\vec{r})=R \exp [-|\vec r|^2/D_0^2]$ 
is diaphragm function 
corresponding to transversely inhomogeneous linear losses, 
key parameter for comparison with 
experiments \cite{Chen:2001} was $D_0 \sim 50 - 2000 \mu m$ ,  
$G_n(\vec{r})=\sigma N_{n}(\vec{r})L_a $ is transversely 
inhomogeneous linear gain, 
$\sigma$ is simulated emission 
cross-section, $N_n(\vec r)$ is density  
of resonant ions per unit volume, $\delta \omega$
is detuning from gain line 
center, $T_2$ and $T_1$ are transverse and longitudinal 
relaxation times correspondingly, $\delta E (\vec r)$ is 
random noise due to spontaneous emission 
term emulated as multimode random process 
\cite{Okulov:1991}, $\delta N (\vec r)$ is fluctuating 
part of optical pump also defined as spatially smooth 
multimode random process,
$N_f=D_0^2/(\lambda L_r)$ is Fresnel 
number, $\lambda $ is lasing wavelength. 
The convergence rate to equilibrium solutions
(stationary  eigenmodes) for dicrete time step 
$\delta t=2 L_r n /c$ proved to be 20 - 100 iterates 
\cite{Okulov:1994}.

For modeling of nonstationary behavior approximately 
the same amount 
of transient iterates had been required \cite{Okulov:1993} to 
attain a realistic oscillation regime with smooth
spatial field profile $E_n(\vec r)$.
The most reliable convergence to realistic solutions had
 been observed for relaxation oscillations (fig.1,2). For low
 Fresnel  number $1<N_f<10$ the transient behavior of 
numerical scheme had been pretty good
 realistic from the very first iterates started from completely
 random  transverse field $E_n(\vec r)$ taken as multimode 
random process with $N_m$ spatial harmonics with random 
phases $\delta \phi_n$ \cite{Okulov:1991}:

\begin{eqnarray}
\label {random initial}
\delta E_n (\vec r)= {N_m}^{-1}\sum_m^{N_m} |A_1 + A_2 + ...
+ A_m +...|^2{\:}{\:}{\:} ; 
&& \nonumber \\
A_m= \delta A_0 \exp [i  \vec k_m \cdot {\vec r}
+ \delta \phi_n ],
\end{eqnarray}
  
 Numerical arrays $E_n (\vec r)$ (complex) and 
$N_n (\vec r)$ (real)
were composed of [128,128] and [256,256] 
points to facilitate the usage of fast Fourier transform 
\cite {Okulov:1993} optimized for standard Fortran compiler.
In most realizations at $1<N_f<10$ the 
transverse field distribution $E_n(\vec r)$ had been almost 
identical to fundamental $TEM_{00}$ gaussian mode 
$E_n(\vec r) \sim \exp [-|\vec r|^2 / D_0^2]$ with slightly 
different $D_0$. 

\begin{figure} 
{ \includegraphics[width=4.5 cm]
{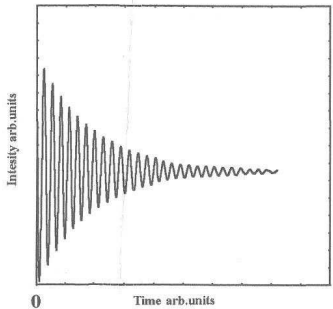}}
\caption{Temporal evolution of intensity $| E_n(\vec r)|^2$ under stepwise
gain switching.} 
\label{fig.1}
\end{figure}

\begin{figure} 
\center
{ \includegraphics[width=4.5 cm]
{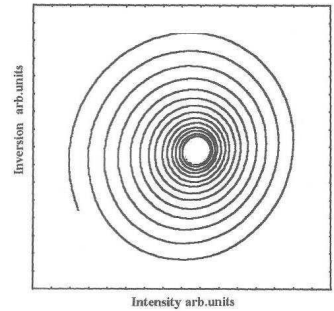}}
\caption{Temporal evolution in phase-space 
of intensity $| E_n(\vec r)|^2$ and inversion $N_n$ under
 stepwise
gain switching. } 
\label{fig.2}
\end{figure}

Increase of Fresnel number have led to perpetual
increase of complexity of output pattern. In 
most cases for $10 < N_f< 50$ 
the stationary output resulted in selection of Hermite- 
Gaussian and Laguerre-Gaussian 
functions. All these low-order eigenfunctions of $\bf 2D$ 
of harmonic oscillator \cite{Chen:2002,Chen:2018} were also 
subjected to relaxation oscillations (fig.3) due to 
intermediate value of cavity lifetime $\tau_c$: 
\begin{equation}
\label {lasing conditions}
 T_1 < \tau_c  <  T_2 ,{\:}{\:} \tau_c = 2 L_r n/ 
(c (1-R)) ,
\end{equation}
The characteristic frequency $\omega_{rel}$ for 
observed relaxation oscillations was close to:
\begin{equation}
\label {relosc}
\omega_{rel} \sim \sqrt { T_1 \tau_c} \sqrt{G-1}; {\:}{\:}{\:} 
G=\sigma N_0({\vec r} = 0).
\end{equation}
In a parameter region between $8<N_f<50$ and $1.5 <G-1< 3$ 
a certain realizations on initial conditions 
$\delta E_n(\vec r)$
have led to localized excitations close to dissipative solitons 
\cite {Okulov:1988,Okulov:2000} of $2D$ hyperbolic secant profile 
\cite {Okulov:2004ksf, Malomed:2002}.
\section{Singular lattices and their vibrations}
The stationary patterns at higher Fresnel numbers 
$10^2 < N_f< 10^3$ demonstrated 
a remarkable phase locking phenomena. The synchronized 
modes of empty resonator 
form an almost perfect periodic lattice of dark spots 
\cite{Okulov:2009_vav}. Each dark spot is 
surrounded by a nearly circular flow resembling isolated
optical vortex alike Laguerre-Gaussian beam 
\cite{Okulov:2012josa,Lembessis:2016_jpb}. 
The phase of optical field 
is a precise indicator of electromagnetic energy circulation
\cite{Lembessis:2016}. When phase is increased from $0$ 
to $2\pi$ along right-handed circular path around amplitude 
$E_n(\vec r)$ zeros, the azimuthal flow of probability 
to find photon in a given volume is accompanied 
by right-handed azimuthal 
component of Pointing vector 
$\vec S=\epsilon_0 c^2 \vec E \times \vec H$ 
\cite {Okulov:2009_vav}. The swirling of electromagnetic 
energy is counter-clockwise in opposite case of 
left-handed probability flow(fig.5). 
A part of numerically obtained patterns is composed from 
zeros of amplitudes equispaced in a checkerboard geometry 
with alternating circulations as is shown 
in \cite {Okulov:2004}. The 
macroscopic photons wavefunction $\Psi \sim E_n(\vec r)$ 
\cite{Okulov:2012josa} which is visualized 
by electric field amplitude $E_n(\vec r)$, demonstrates the
regular lattices of field zeros 
with counter-rotating whirls around [fig.6]. 
In contrast to optical 
speckle patterns \cite{Okulov:2009} where vortex-antivortex 
pairs are distributed chaotically in space 
the lattice of whirls 
in wide-aperture lasers is periodic in space [fig.5]. The far 
field pattern consists of four lobes whose separation is 
defined by wavelength $\lambda$ and 
period $p$ of the near field pattern \cite{Okulov:2008}.  
\begin{figure} 
\center
{ \includegraphics[width=5.0 cm]
{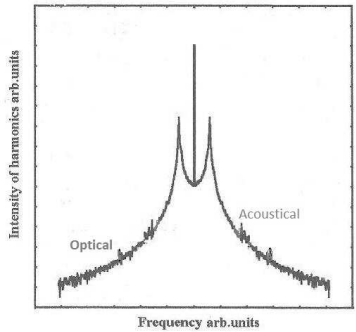}}
\caption{The power spectrum 
of wide area (high Fresnel number $100 <N_F < 1000$) stationary
emission pertubed by spontaneous noise added at each round
trip $n$. Noteworthy the collective excitations spikes
 interpreted 
as acoustical and optical vibrations of vortex lattice of 
9x9 whirls . }
\label{fig.3}
\end{figure}

\begin{figure} 
\center
{ \includegraphics[width=5.0 cm]
{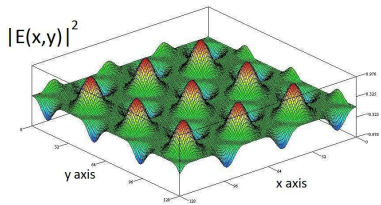}}
\caption{The stationary transverse intensity distribution 
$| E_n(\vec r)|^2$ 
of wide area laser stationary
emission with high Fresnel number $100 <N_F < 1000$ .}
\label{fig.4}
\end{figure}
The other set of numerical realizations appears to be 
identical to Abrikosov lattice obtained analytically 
$\Psi \sim \vartheta_3 [\sqrt{2 \pi} i \kappa (x+iy);1]$ 
\cite {{Abrikosov:1957}},where $\kappa$ is ratio 
of coherence length to penetration depth, $\vartheta_
3 $ is Jacobi theta function (fig.4) Exactly  
as in the early models of type-II superconductors placed in
external magnetic field \cite{Abrikosov:1957} the vortices 
at zeros of amplitudes have co-directed angular momenta 
but the field of velocities is continuous as well 
due to rotation 
around bright spots. The energy flow around maxima  
of intensity is opposite to flow around zeros. Both types of 
vortex lattices obtained numerically have remarkable square
symmetry found yet in \cite{Abrikosov:1957}. Noteworthy 
experimentally observed lattices in superconductors of the second 
type have triangular symmetry.  
 
The noise sources due to 
spontaneous emission $\delta E_n (\vec r)$ 
and fluctuations of optical pump $\delta N_n (\vec r)$ 
have led to vibrations of vortex lattices. To detect vibrations 
the power spectrum 
$I(\Omega)= FFT [|E_n(\vec r)|^2]$ had been recorded at several 
points $\vec r$ of transverse section. In the 
stationary regime with 
well formed vortex lattice the spectrum $I(\Omega)$ 
proved to be independent of location $\vec r$. The large 
maximum near frequency of relaxation oscillations had been 
recorded in all oscillation regimes and for all transverse 
spatial distributions. The remarkable feature of obtained 
patterns are small maxima of power spectrum 
beyond relaxation frequency. The early numerical study 
reported this spikes at power spectrum 
as collective vibrations of vortex 
lattice \cite {Staliunas:1995}. The mechanism responsible for 
lattices vibrations had been reported as 
vortex precession around equilibrium 
position. In our numerical model the observed 
vortex lattices of 
4x4, 7x7 and 9x9 whirls were practically unmovable at 
a long discrete time $n$ intervals. Nevertheless the 
sufficiently large vortex lattice composed of 9x9 whirls 
produced numerical noise spectrum with a small but clearly 
visible collective excitation spikes (fig.3). As is known 
from lattice vibration theory there exist two basic 
types of collective lattice vibrations with a different 
dispersion laws. The acoustic mode 
corresponds to running waves whose group velocity 
tends to zero at the edge of the first Brillouin zone.
The optical mode of vibrations 
possesses the higher frequencies with 
significantly smaller group velocities. The discrete time 
$n$ to record this this spectrum $I(\Omega)$ was about 
500 time steps (iterates of maps(1-4)). The runtime took 
for about a one minute on $dual$ $core$ $1.86 Ghz$ processor 
per each FFT accelerated 
iteration of discrete master equations (1-4).  
 
\begin{figure}
\center
{ \includegraphics[width=4.0 cm]
{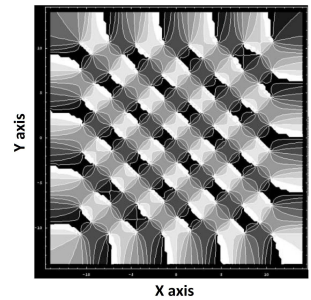}}
\caption{The grey scale stationary transverse 
phase distribution 
$ arg [ E_n(\vec r)|]$ 
of wide area (high Fresnel number $100 <N_f < 1000$) 
laser . Black color corresponds to phase 
$ arg [ E_n(\vec r)|]=0$ which gradually 
increases to $ arg [ E_n(\vec r)|]=2 \pi$ 
in circumvention around vortex cores.}
\label{fig.5}
\end{figure}

\begin{figure}
\center
{ \includegraphics[width=4.0 cm]
{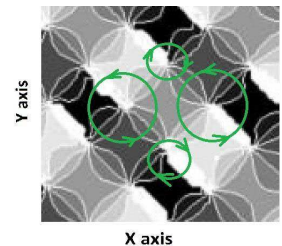}}
\caption{The enlarged elementary cell
grey scale transverse 
phase distribution 
$ arg [ E_n(\vec r)|]$ 
of wide area (high Fresnel number $100 <N_f < 1000$) 
laser nonstationary emission. }
\label{fig.6}
\end{figure}

\section{Conclusion}

The computational model for eigenmodes $E_n(\vec r)$
of microchip laser resonator is 
composed of convolution integral with nonlinear kernel
evaluated via Fast-Fourier transform  
\cite{Okulov:1988, Okulov:1994} and 
the relaxation oscillator $N_n(\vec r)$ 
for gain medium coupled with
wavefunction $E_n(\vec r)$. 
The regular square lattices (fig.4,5)
appear in a well defined range of key system parameters namely 
laser gain and Fresnel number.  
The inherent feature of these nonstationary vortex 
patterns are vibrations of observed lattices which are known as 
optical and acoustical ones (fig.3).


\begin{thebibliography}{99}

\bibitem{Cross:1993}{ Cross M. C. and Hohenberg P. C. } 1993 
Pattern formation outside of equilibrium 
{\it Rev. Mod. Phys.}, {\bf 65}, 851 

\bibitem{Okulov:1986} {Okulov A.Yu.and Oraevsky A.N.} 1986
Spatio-temporal behavior of a light pulse in nonlinear 
nondispersive media {\it J.Opt.Soc.Am} B {\bf 3} 741

\bibitem{Chen:2001}{ Chen Y. F. and Lan Y. P.} 2001 
Transverse pattern formation of optical vortices 
in a microchip laser with a large Fresnel number 
{\it Phys. Rev.} A {\bf 65}, 013802 

\bibitem{Staliunas:1995}{K.Staliunas and C.O.Weiss} 1995 
Nonstationary vortex lattices in large-aperture class B 
 lasers {\it J.Opt.Soc.Am} B {\bf 12}, 1142 

\bibitem{Okulov:2008}{Okulov A.Yu.} 2008 
3D-vortex labyrinths 
in the near field of solid-state microchip laser 
{\it J.Mod.Opt.} {\bf 55} , 241-259

\bibitem{Francois:2014}{Francois N., Xia H., Punzmann H., 
 Ramsden S. and Shats M.,} 2014 
Three-dimensional fluid motion in Faraday waves: 
creation of vorticity and generation of two-dimensional
 turbulence {\it Phys. Rev.} X {\bf 4}, 021021 

\bibitem{Chen:2003}{Chen Y. F. and Lan Y. P.} 2003 
Observation of transverse patterns in an 
isotropic microchip laser 
{\it Phys. Rev.} A, {\bf 67(4)}, 043814

\bibitem{Okulov:2004}{Okulov A.Yu.} 2004
3D-configuration of the vortex lattices in microchip laser
cavity 
QCMC-2004, {\it AIP Conference Proceedings} {\bf 734}, p.366 

\bibitem{Okulov:2009_vav}
{Okulov A.Yu.} 2009 Vortex-antivortex wavefunction 
of a degenerate quantum gas {\it Laser Physics} 
{\bf 19}, 1796-1803 

\bibitem{Malomed:2018}{Burlak G. and Malomed B.A.} 2018 
Interactions of three-dimensional solitons in the 
cubic-quintic model 
{\it Chaos} {\bf 28}, 063121
 
\bibitem{Okulov:1988}{Okulov A.Yu.and Oraevsky A.N.}1988 
Spatiotemporal dynamics of a wave packet in nonlinear 
medium and discrete maps, 
{\it Proceedings Lebedev Physics Institute (in Russian) 
N.G.Basov ed., Nauka, Moscow} {\bf 187}, 202-222

\bibitem{Khazanov:2013}{ Vadimova O.L., Mukhin I.B., 
Kuznetsov I.I.,Palashov O.V.,Perevezentsev E.A. and 
Khazanov E. A.} 2013 Calculation of the gain coefficient in
 cryogenically cooled Yb : YAG disks at high heat 
generation rates {\it Quantum Electronics} {\bf 43}, 
n.3, 201-206

\bibitem{Okulov:1991}{Okulov A.Yu.}1991 The effect of 
roughness of optical elements on the transverse 
structure of a light field in a nonlinear Talbot cavity 
{\it J.Mod.Opt.} {\bf 38}, n.10, 1887

\bibitem{Okulov:1994}{Okulov A.Yu.}1994
On correlations between cavity mode and inversion profile 
in a solid-state microchip laser
{\it Optics and Spectroscopy} {\bf 77}, n.6, 888-892  

\bibitem{Okulov:1993}
{Okulov A.Yu.} 1993 Scaling of diode-array-pumped solid-state 
lasers via self-imaging 
{\it Opt.Comm.} {\bf 99}, p.350-354

\bibitem{Chen:2002}{Chen Y. F.and Lan Y. P.,} 2002 
Observation of laser transverse modes analogous to a SU(2)
wave packet of a
quantum harmonic oscillator 
{\it Phys. Rev.} A{\bf 66(5)}, 053812

\bibitem{Chen:2018}{Chen Y. F. , Hsieh Y. H., and Huang K. F.} 
2018 Originating an integral formula and using the 
quantum Fourier
transform to decompose the Hermite-Laguerre-Gaussian modes
into elliptical orbital modes 
{\it OSA Continuum}, {\bf 1}, Issue 2, pp. 744-754

\bibitem{Okulov:2000}{Okulov A.Yu.} 2000 
Spatial soliton laser: geometry and stability 
{\it Optics and Spectroscopy} {\bf 89}, 145-147
 
\bibitem{Okulov:2004ksf}{Okulov A.Yu.} 2004 
Exact theory of localized and periodic structures in an
active optical resonator
{\it Bulletin of Lebedev Physical institute}, n.10

\bibitem{Malomed:2002} {Malomed B.A. and Towers I.} 2002
Stable (2+ 1)-dimensional solitons in a layered medium with
 sign-alternating Kerr nonlinearity {\it J.Opt.Soc.Am} B
 {\bf 19} 537-543


\bibitem{Okulov:2012josa}{Okulov A.Yu.} 2012
Rotational Doppler shift of the 
phase-conjugated photons, {\it J.Opt.Soc.Am} B 
{\bf 29}, 714-718 

\bibitem {Lembessis:2016_jpb} { Rsheed A. Al, Lyras A., 
Lembessis V. E., Aldossary O. M.} 2016
Guiding of atoms in helical optical potential
structures {\it  J. Phys.} B {\bf 49} 125002

\bibitem {Lembessis:2016} { Rsheed A. Al, Lyras A., 
Lembessis V. E., Alqarni A., Alshamari S., Siddig A., 
Aldossary O. M.} 2016 
Rotating optical tubes for vertical transport of atoms
{\it Phys. Rev.} A {\bf 94}, 063423

\bibitem{Okulov:2009}
{Okulov A.Yu.} 2009 Twisted speckle entities inside wavefront
 reversal mirrors
{\it Phys. Rev.} A {\bf 80}, 013837

\bibitem{Abrikosov:1957}{Abrikosov A.A.} 1957 
On the magnetic properties of superconductors 
of the second group 
{\it JETP} {\bf 5} 1174

\end{thebibliography}
\end{document}